\documentclass[aps,prl,reprint]{revtex4-1}

\usepackage{blindtext}
\usepackage[export]{adjustbox}
\usepackage{graphicx,epsfig,epstopdf}
\usepackage{amsmath}
\usepackage{color}
\usepackage{subfigure}
\usepackage{array}
\usepackage{tabularx}
\usepackage{multirow}
\newcolumntype{L}[1]{>{\raggedright\arraybackslash}p{#1}}
\newcolumntype{C}[1]{>{\centering\arraybackslash}p{#1}}
\newcolumntype{M}[1]{>{\centering\arraybackslash}m{#1}}

\def\_#1{{\bf #1}}
\def\.{\cdot}

\def\Im{{\rm Im\mit}}

\begin{document}

	\title{Bianisotropic metasurfaces for scattering-free manipulation of acoustic wavefronts}
	
	\author{Junfei Li$^{1,\star}$, Chen Shen$^{1,\star}$, Ana~D\'{i}az-Rubio$^{2}$, Sergei~Tretyakov$^{2}$,   and Steven Cummer$^1$}
	
	\affiliation{$^1$Department of Electrical and Computer Engineering, Duke University, Durham, North Carolina 27708, USA\\
		$^2$Department of Electronics and Nanoengineering, Aalto University, P.~O.~Box~15500, FI-00076 Aalto, Finland\\
        $^\star$ J.L. and C.S. contributed equally to this work}

\begin{abstract} 
Recent advances in gradient metasurfaces have shown that only by locally controlling the bianisotropic response of the constituent cells can one ensure full control of refraction, i.e., arbitrarily modify the direction of the incident waves without producing scattering into unwanted directions.  In this work, we propose and experimentally verify the use of a new acoustic cell architecture that provides enough degrees of freedom to fully control the bianisotropic response and minimizes the implementation losses produced by resonant elements. The versatility of the approach is shown through the design of three different anomalous refractive metasurfaces capable of redirecting a normally incident plane wave to 60, 70 and 80 degrees on transmission. The efficiency of the bianisotropic designs is over 90\%, much higher than the corresponding designs based on the conventional generalized Snell’s law  (81\%, 58\%, and 35\%). The proposed strategy opens a new way of designing practical and highly efficient bianisotropic metasurfaces for different functionalities, enabling nearly ideal control over the acoustic energy flow through thin metasurfaces.
\end{abstract}

\maketitle
\section{Introduction}
	
The ability to fully control the behavior of classical waves (e.g., electromagnetic and acoustic waves) has long been desired and is at present a highly active research area. Among numerous routes to achieve this, metamaterials have served as a primary approach in recent years 
\cite{cummer2016controlling,zheludev2012metamaterials}.
The possibilities are enabled, in one approach, by engineering subwavelength structures in the fashion that local resonance dominates and the overall constitutive parameters can take arbitrary values which are not found in nature. In contrast to the volumetric modulation using metamaterials,  two-dimensional arrangements of subwavelength cells offer an alternative solution of molding wave propagation within a planar or nearly flat geometry. These two-dimensional patterned surfaces, or “metasurfaces”, have opened up unprecedented possibilities for controlling waves at will \cite{chen2016review,glybovski2016metasurfaces}.
One of the most attractive aspects of metasurfaces is the ability to engineer the scattered wavefronts by packing phase shifts along the metasurface. These locally non-periodic metasurfaces, also known as gradient metasurfaces, have awakened interest as a possible approach for the design of lenses, beam splitters, and more. \cite{xu2016planar,estakhri2016recent}.

In both the electromagnetic \cite{sun2012gradient,yu2011light,kildishev2013planar} and acoustic \cite{tang2014anomalous,li2014experimental,li2013reflected,zhao2013manipulating,wang2014design,wang2016subwavelength,xie2014wavefront} communities, the conventional gradient metasurface design approach is based on the implementation of a local phase modulation which dictates the behavior of outgoing waves according to the generalized Snell's law (GSL) \cite{li2013reflected}. 
In acoustics, various unit cell topologies have been proposed to achieve a homogenized effective index to control the local transmitted or reflected phase shift, such as labyrinthine cells \cite{tang2014anomalous,li2014experimental,li2013reflected}, spiral cells \cite{wang2014design,wang2016subwavelength,xie2014wavefront}, helical cells \cite{zhu2016implementation}, and pipes with side-loaded resonators \cite{li2015metascreen,li2016theory}, to name a few. Knowledge of gradient index materials has been applied to acoustic devices for different functionalities, such as wavefront manipulation \cite{tang2014anomalous,li2014experimental,li2013reflected,zhao2013manipulating,wang2014design,wang2016subwavelength,xie2014wavefront}, sound absorption \cite{xie2014wavefront,li2016theory,li2016sound}, asymmetric transmission \cite{shen2016asymmetric} and carpet cloaking \cite{zhang2013ultrathin,esfahlani2016acoustic}. However, the efficiency of phase shift devices is fundamentally restricted by the scattering into unwanted directions, which hinders their use for aberration-free applications. The origin of the problem is attributed to the local reflection produced by the individual unit cells and, to enable better performance, many approaches have been applied to improve the transmission of the unit cells, such as making helical cells \cite{zhu2016implementation}, tapered spiral cells \cite{xie2013tapered}, changing the geometry of cell apertures \cite{xie2013measurement,memoli2017metamaterial}, or filling the channel with light materials \cite{jahdali2016high}. 
    
\begin{figure*}
		\subfigure[]{\includegraphics[width=0.6\linewidth]{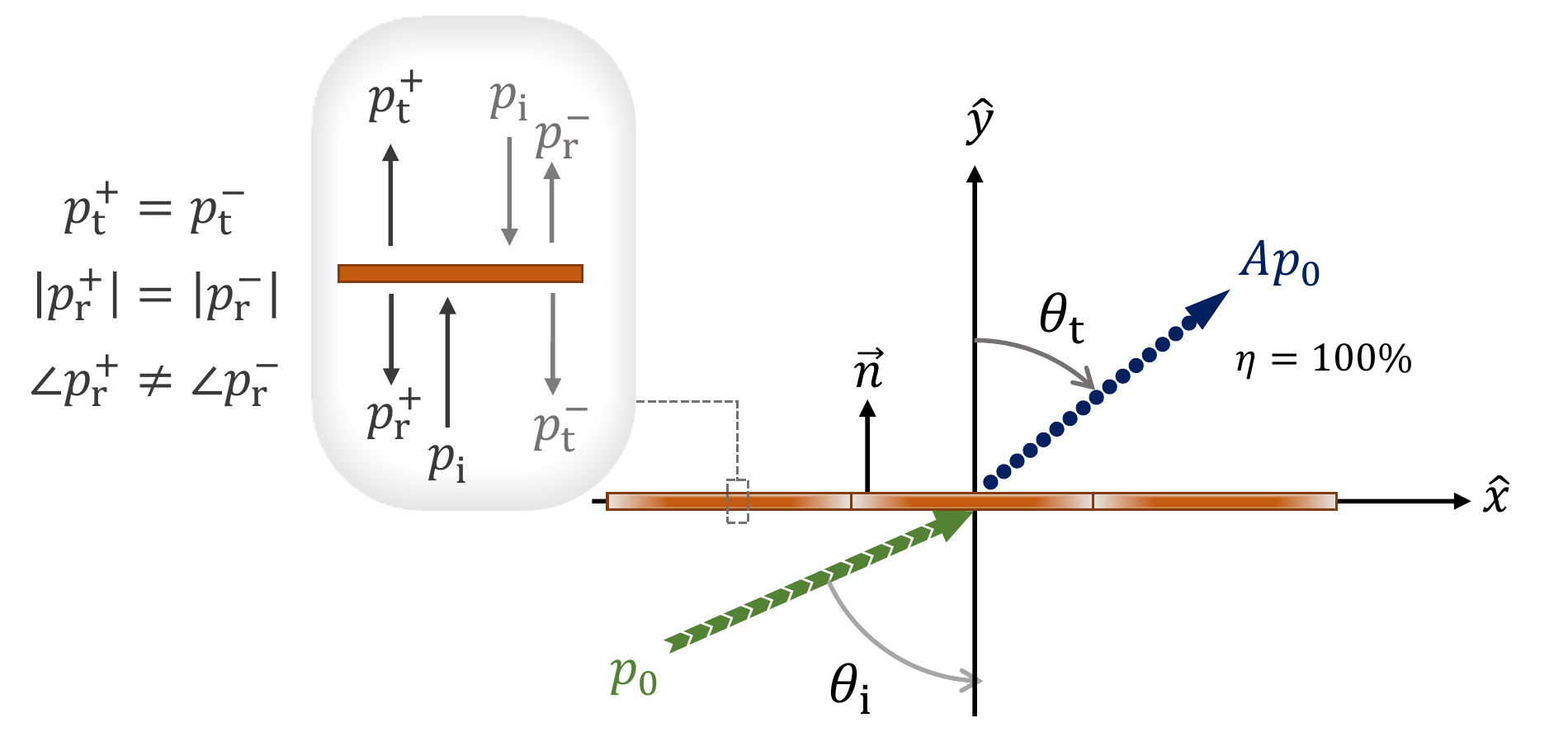}\label{fig:Fig1A}}
		\subfigure[]{\includegraphics[width=0.35\linewidth]{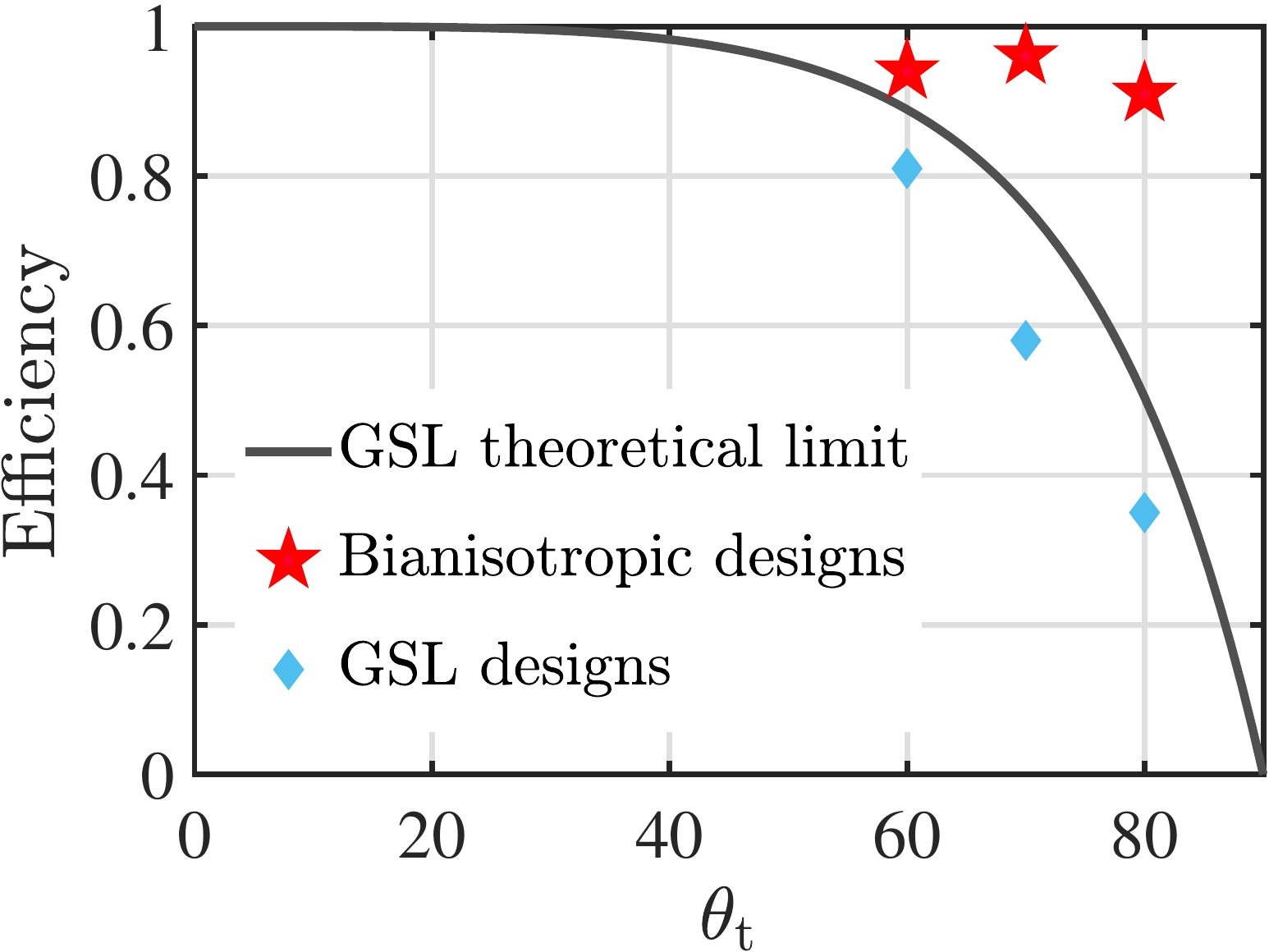}\label{fig:Fig1B}}
		\caption{(a) Illustration of the desired performance of a perfect anomalous refractive metasurface. All the energy is pointing out toward the desired direction with no parasitic scattering. The inset shows the bianisotropic response of the elements of the metasurface, i.e. the asymmetric response for incident waves from opposite directions (the phase of the reflected wave are different). (b) Comparison of the efficiency for anomalous transmission metasurfaces. Bianisotropic designs show great advancement especially for large deflection angles. }
\label{fig:Fig1}
\end{figure*}    
   
Recent work has shown that a local phase gradient alone cannot provide full control over the scattered wave and therefore it is not possible to completely control the scattered field \cite{diaz2017acoustic,wong2016reflectionless,asadchy2016perfect,epstein2016synthesis,diaz2017from,estakhri2016wave,asadchy2017eliminating}. Let us now consider the anomalous refraction as an example, which is the simplest functionality offered by gradient index metasurfaces in transmission [see Fig.\ref{fig:Fig1A}].  
For an optimal performance, the metasurface has to transmit all the illuminating energy into another arbitrary direction. 
As it was pointed out for electromagnetic and acoustic waves, there is a fundamental limitation associated with all conventional designs based on the generalized Snell's law that originates in the impedance mismatch between incident and refracted waves. Figure \ref{fig:Fig1B} shows the theoretical efficiency for a metasurface illuminated normally as a function of the angle of refraction. This limitation is inherent to the design approach and it does not depend on the microscopic topology used in an actual implementation.  In order to overcome the problem, one has to ensure the control of not only the phase gradient along the metasurfaces but also the impedance matching between the incident and the desired scattered waves. 

Rigorous analysis of the problem has shown that
the macroscopic impedance matching required for theoretically perfect anomalous refraction of plane waves can be realized if the metasurface exhibits bianisotropy: magneto-electric  coupling  for electromagnetic metasurfaces  \cite{wong2016reflectionless,asadchy2016perfect,epstein2016synthesis} and Willis coupling for the acoustic counterpart \cite{diaz2017acoustic}.  The bianisotropic response can be implemented by asymmetric unit cells, where the scattered fields are different depending on the direction of illumination. For electromagnetic metasurfaces, typical solutions are based on a cascade of three impedance layers  where, by independently controlling the impedance of each layer, the asymmetric response can be fully controlled \cite{lavigne2017refracting,chen2017experimental}. These structures have been numerically and experimentally verified.  

In acoustics, however, practical design or experimental realization has remained scarce. Although bianisotropy in acoustics, also referred to as Willis coupling \cite{willis1981variational,sieck2017origins} in elastodynamics, has been reported recently in a single cell \cite{koo2016acoustic,muhlestein2017experimental}, the integration of bianisotropy into a macroscopic acoustic metasurface for perfect wavefront modulation with controlled asymmetric response is not reported. Such metasurfaces require a control over the amplitudes and phases of the scattered fields which is not offered by previous designs with symmetric inclusions. An successful approach was theoretically proposed by using three membranes which allows full control of the bianisotropic response \cite{diaz2017acoustic}. However, the surface tension and uniformity of the membranes, etc. are extremely difficult to control, and it is questionable whether this design can be adopted in practice.
    
Recent electromagnetic and acoustic studies \cite{wong2016reflectionless,epstein2016synthesis,lavigne2017refracting,chen2017experimental,diaz2017acoustic}, have shown that full control of the asymmetric response requires at least three degrees of freedom in the design of the particles, which can be obtained with a cascade of three independent resonators. Although this topology satisfies the minimum requirements for obtaining arbitrary bianisotropic response, the structures become resonant to obtain extreme values of bianisotropy and will induce a great amount of losses inside the structures.
	
In this work, a versatile platform for bianisotropic metasurfaces based on the use of four independent resonators is developed. To bring about feasible designs for actual implementations, we chose Helmholtz resonators which can be easily controlled by changing the geometrical dimensions. The validity of the proposed bianisotropic particle is tested with the design of three different anomalous refractive metasurfaces able to redirect a normally impinging plane wave into 60, 70, and 80 degrees. In addition, we experimentally characterize the first bianisotropic gradient metasurface for perfect acoustic anomalous refraction. 
	
\section{Results}
 \subsection{Non-resonant bianisotropic acoustic cells}

\begin{figure*}
\minipage{0.45\textwidth}
		\subfigure[]{\includegraphics[width=1\linewidth]{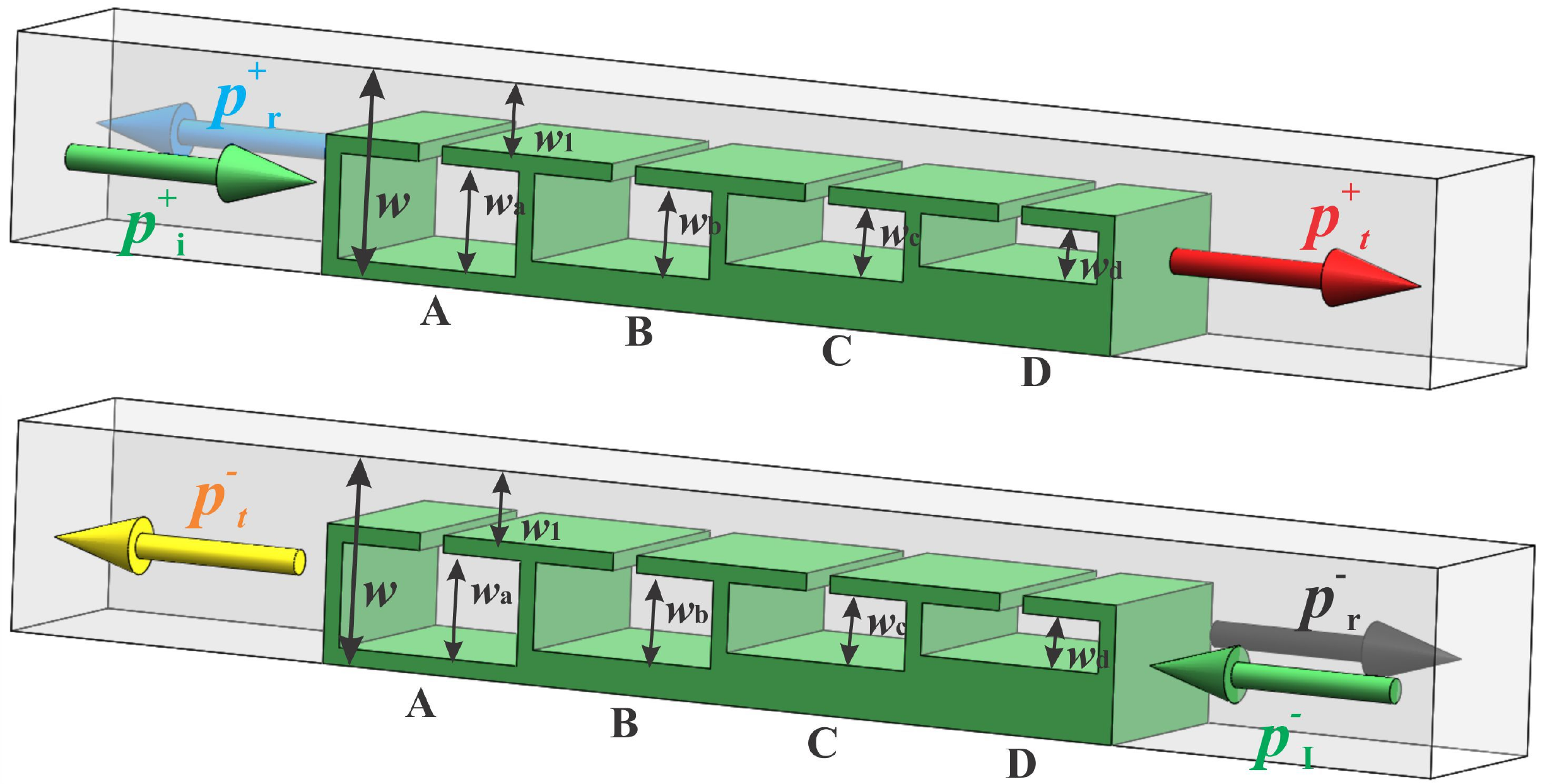}\label{fig:Fig2A}}
\endminipage\hfill
\minipage{0.5\textwidth}
\subfigure{\includegraphics[width=0.95\linewidth,left]{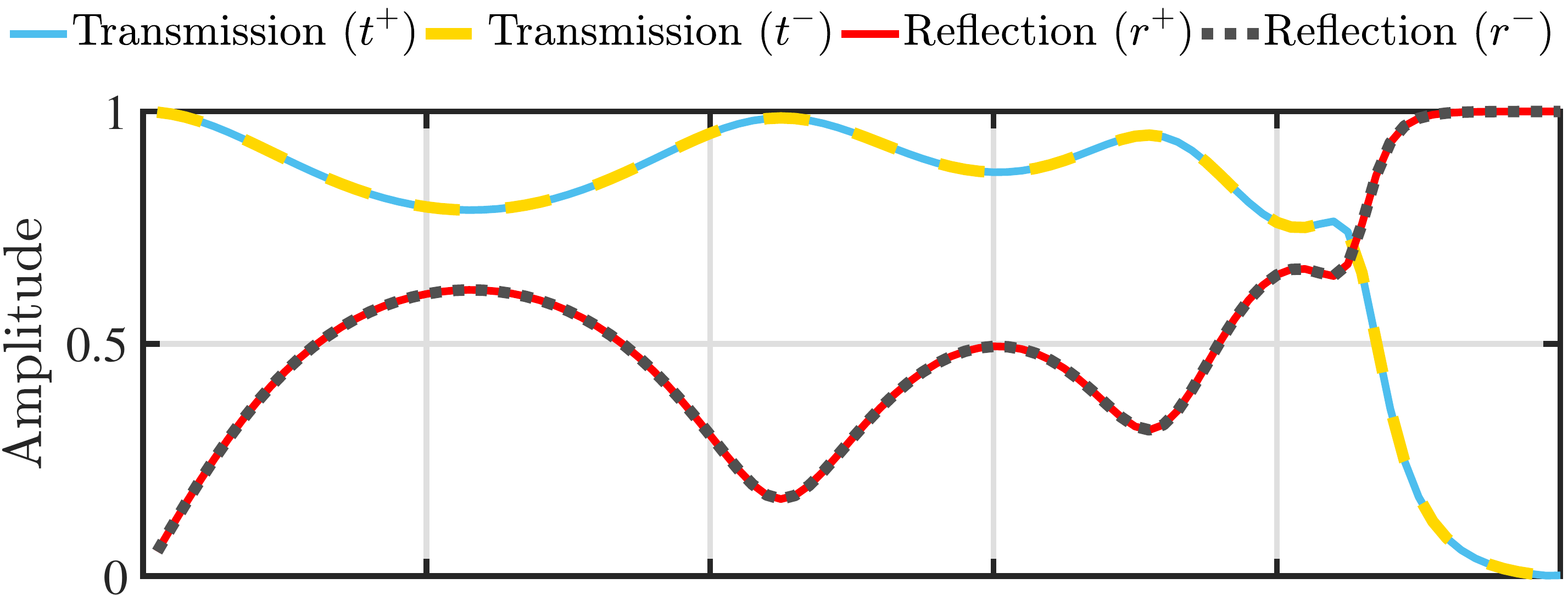}\noindent\label{fig:Fig2B}}
\subfigure{\includegraphics[width=0.95\linewidth,left]{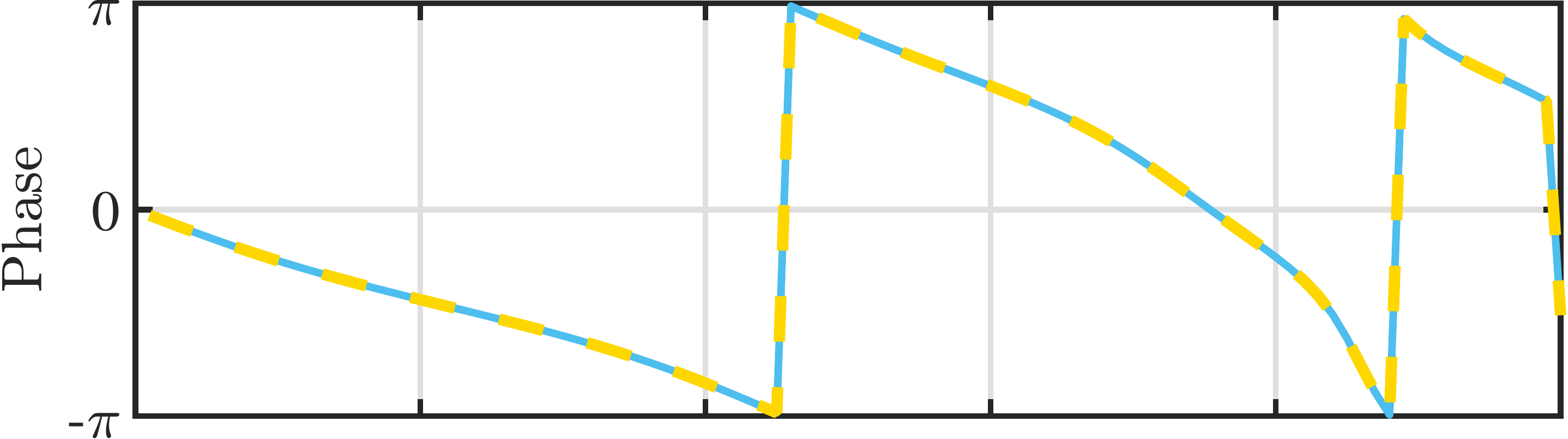}\label{fig:Fig2C}}
\subfigure[]{\includegraphics[width=0.99\linewidth,left]{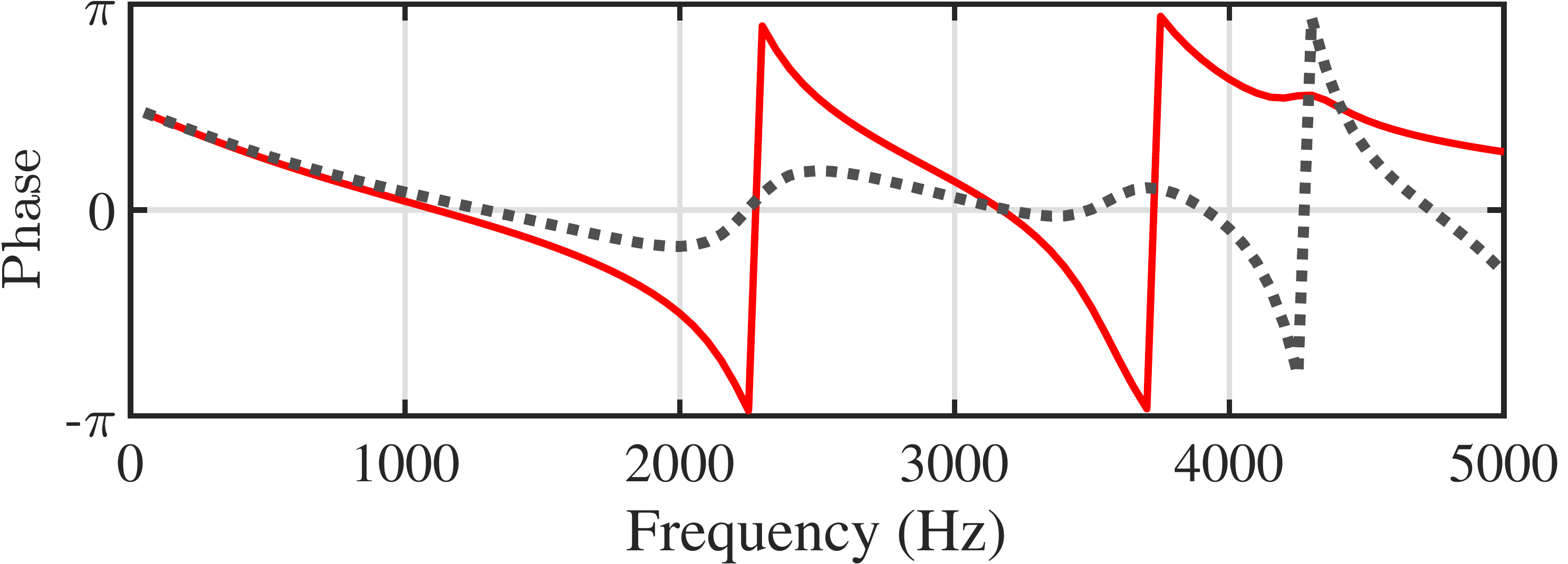}\label{fig:Fig2D}}
\endminipage\hfill
		\caption{Study of a bianisotropic acoustic cell. (a) Geometry of a cell with four side-loaded resonator. The height of the Helmholtz resonators is varied to create different bianisotropic responses. Definition of the forward (+) and backward (-) illuminations. (d) Amplitude and phase of the transmission and reflection coefficients of an arbitrary cell. The dimensions of the cell are: $w=12$ mm, $h_2=1.5$ mm, $w_2=1$ mm, $h_1=1$ mm $w_1=4$ mm, $w_{\rm a}=6$ mm, $w_{\rm b}=5$ mm, $w_{\rm c}=4$ mm, and $w_{\rm d}=3$ mm.}\label{fig:Fig2}
\end{figure*} 
 
The cell architecture that we use to ensure asymmetry, shown in Fig. 2(a), is based on a straight channel with side-loaded resonators. Similar structures with identical resonators have been used to slow down the speed of sound in the channel, controlling the output phase and diffuse pattern \cite{li2015metascreen,li2016theory,jimenez2017metadiffusers} or achieve perfect sound absorption at resonant frequencies \cite{jimenez2016ultra,jimenez2017quasiperfect}. 
However, since the structures are desisnged to be symmetric (i.e., metasurfaces with identical response from opposite directions), they do not exhibit bianisotropy.  In the most general case the relation between the fields at both sides of a lossless array of bianisotropic cells placed at $y=0$  can be expressed as
\begin{equation}
\begin{bmatrix}
p^{\rm +}(x,0)    \\
p^{\rm -}(x,0)
\end{bmatrix}=
\begin{bmatrix}
Z_{11}     & Z_{12}  \\
Z_{21}     & Z_{22}
\end{bmatrix}\begin{bmatrix}
\;\; \hat{n} \cdot\vec{v}^{\rm +}(x,0)    \\
-\hat{n} \cdot\vec{v}^{\rm -}(x,0)
\end{bmatrix}\label{eq:ImpedanceMatrix}
\end{equation}
where $\hat{n}$ is the normal vector of the metasurface, $Z_{ij}$ are the components of the impedance matrix, and the $\pm$ sign refers to the fields at both sides of the metasurface. Note that for such a linear time-invariant system under study, reciprocity requires $Z_{12}=Z_{21}$ and we assume this condition throughout. The cell will have bianisotropic response if $Z_{11}\ne Z_{22}$, and this condition can be satisfied if the acoustic cells has structural asymmetry [see Fig. 2(a)].

From the analysis of the bianisotropic requirements dictated by the impedance matrix, we can see that with the proposed topology  three resonators is the minimum requirement which allows to implement any desired response (Supplementary Note 1). However, to obtain extreme asymmetric response required by some gradient metasurfaces, the resonators will have to work near resonance frequencies and this will make difficult to control their responses. In order to avoid the use of resonant particles which will increase the losses, we propose a four side-loaded resonators particle, as shown in Fig. \ref{fig:Fig2A}. In this structure: the width and height of the neck , $h_2$  and $w_2$, are fixed in the four resonators; the width of the cavities $h_3$ is also fixed; the height of the air channel $w_1$ and the height of the resonators $w_{\rm a}$, $w_{\rm b}$, $w_{\rm c}$ and $w_{\rm d}$ can be varied to control the asymmetry; and the wall thickness of the unit cell $h_1$ is fixed and will be defined with the fabrication limitations.

 \begin{figure*}
		\minipage{0.32\textwidth}
		\subfigure[]{\includegraphics[height=4.5cm,keepaspectratio]{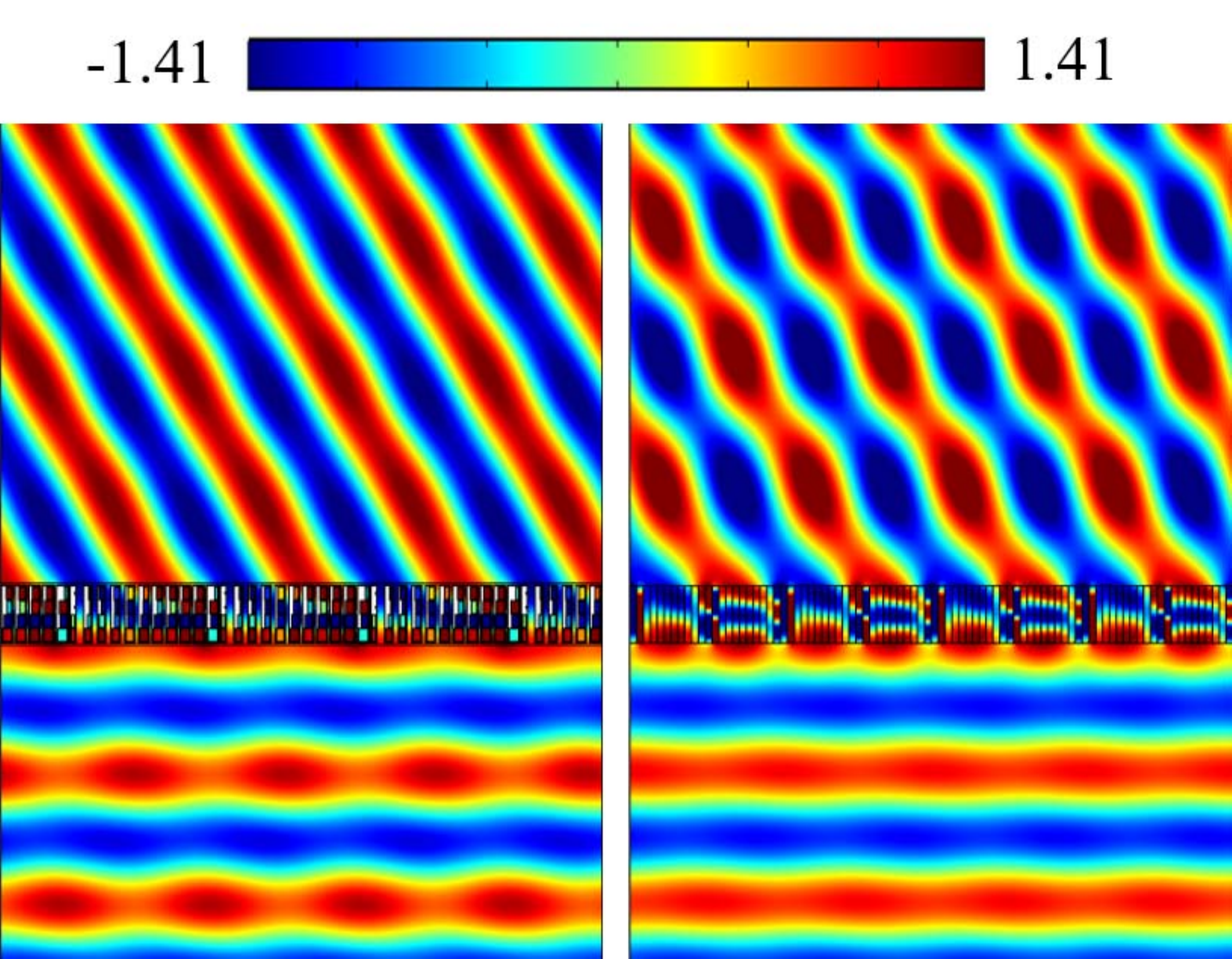}\label{fig:Fig3A}}
		\subfigure[]{\includegraphics[width=1\linewidth]{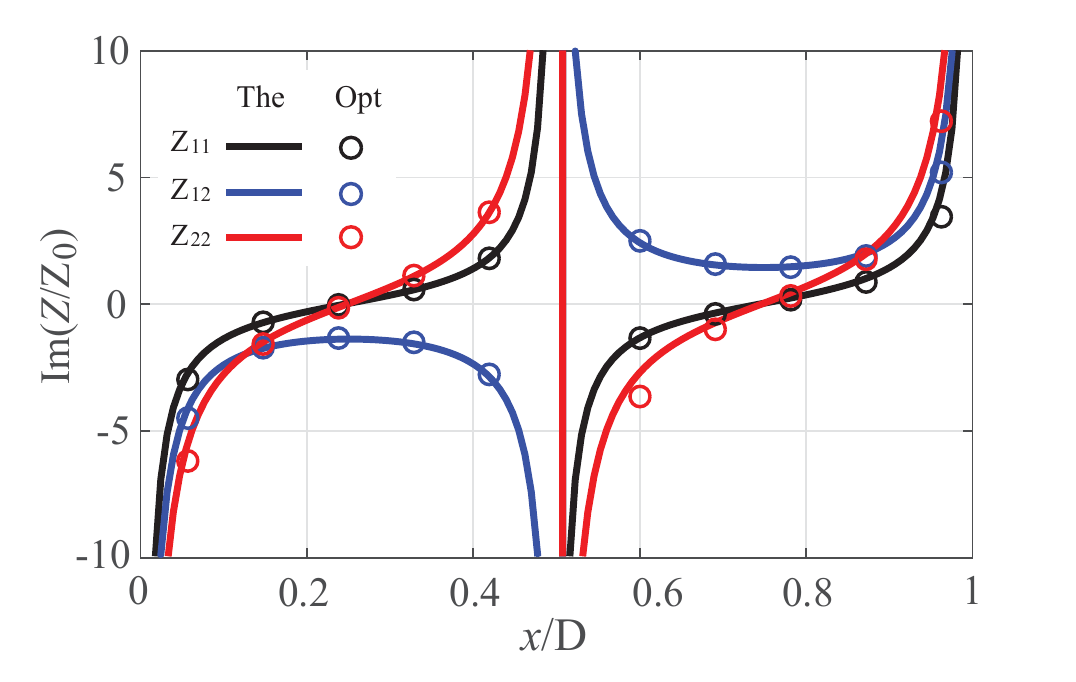}\label{fig:Fig3B}}
		\endminipage\hfill
		\minipage{0.32\textwidth}
		\subfigure[]{\includegraphics[height=4.5cm,keepaspectratio]{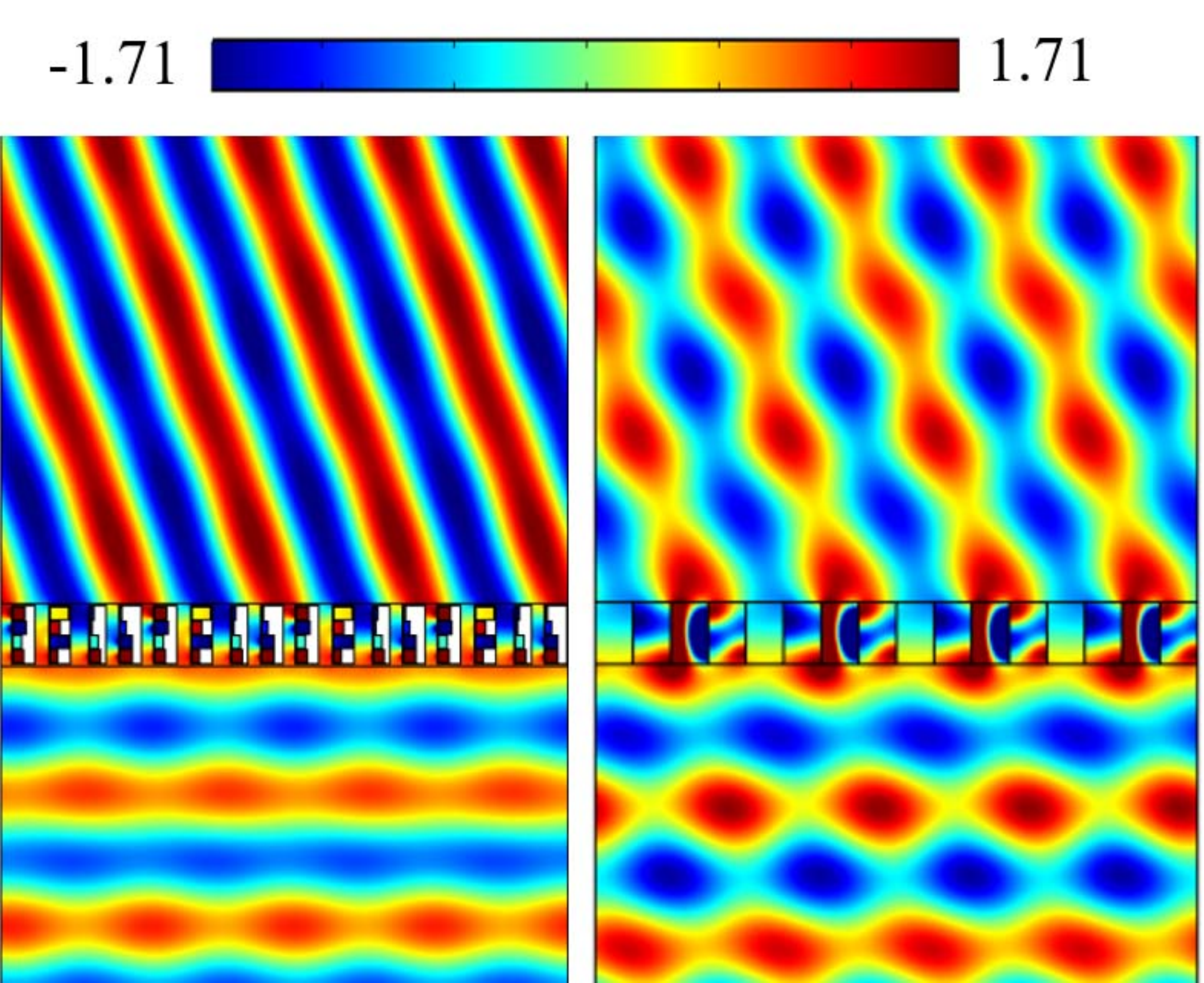}\label{fig:Fig3C}}
		\subfigure[]{\includegraphics[width=1\linewidth]{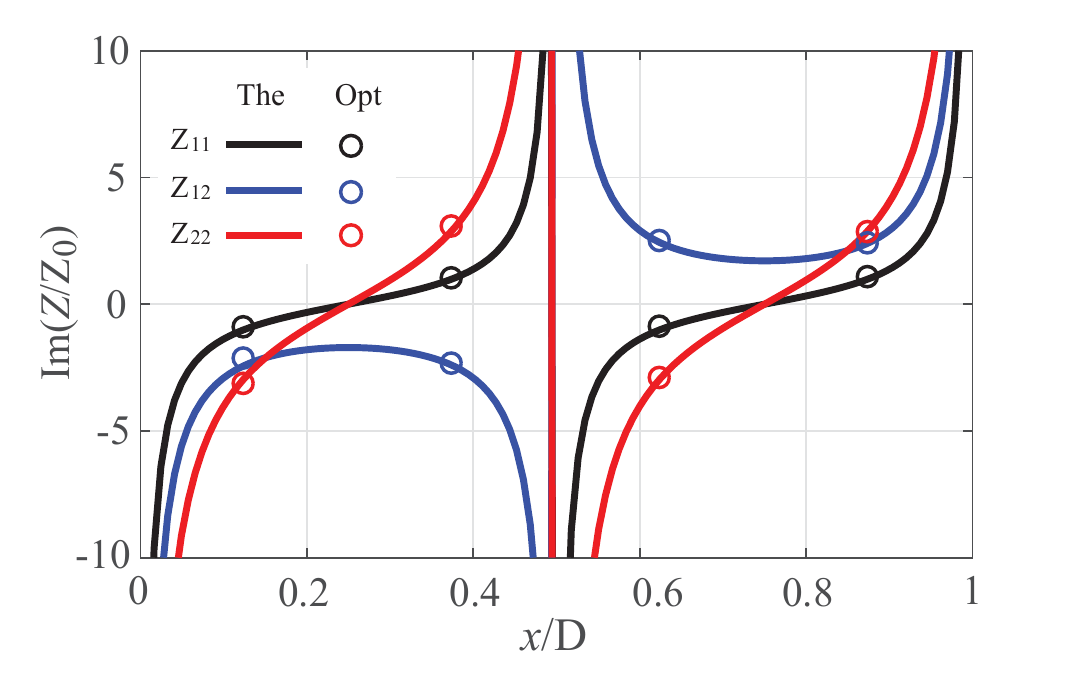}\label{fig:Fig3D}}
		\endminipage\hfill
		\minipage{0.32\textwidth}
		\subfigure[]{\includegraphics[height=4.5cm,keepaspectratio]{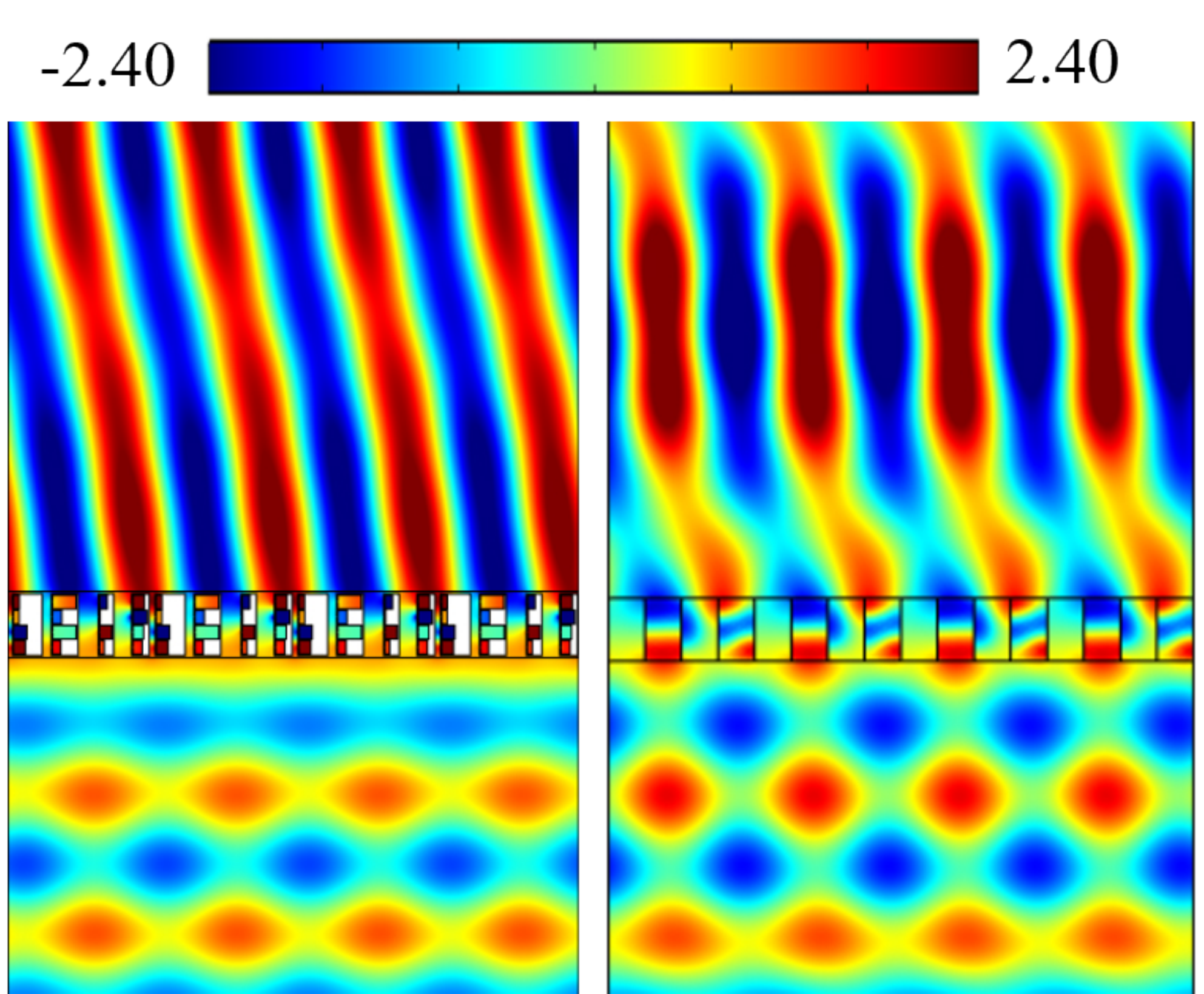}\label{fig:Fig3E}}
		\subfigure[]{\includegraphics[width=1\linewidth]{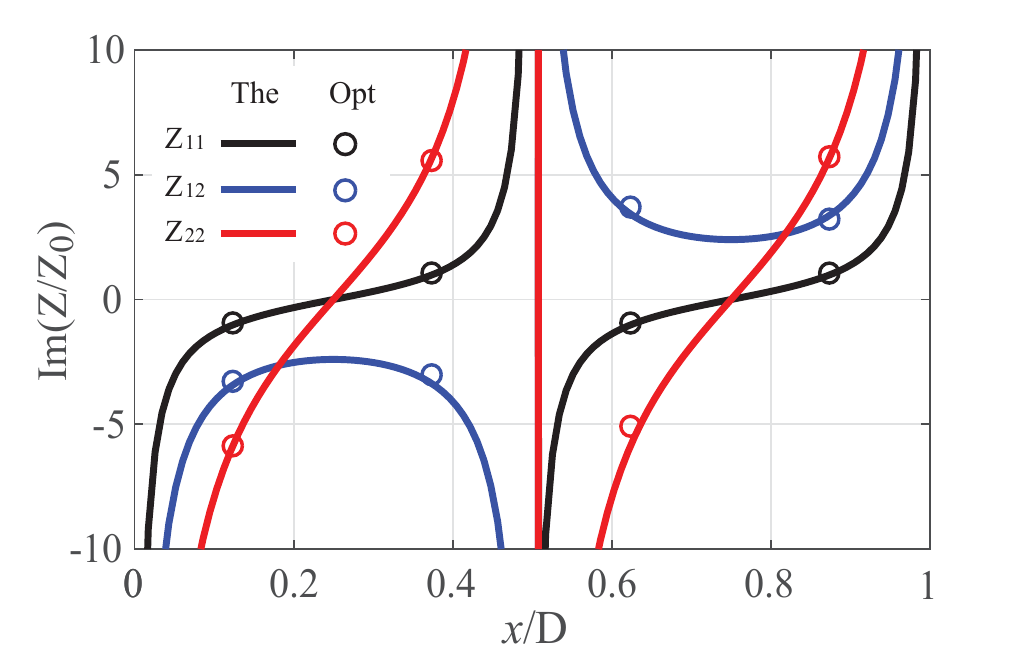}\label{fig:Fig3F}}
		\endminipage\hfill
		\caption{Bianisotropic metasurfaces for scattering-free anomalous refraction.  (b), (d), and (f) represent the impedance matrices profile for $\theta_{\rm i}=0^\circ$ and $\theta_{\rm t}=60^\circ$, $70^\circ$, and $ 80^\circ$. 
 (a), (c), and (d) represent the numerical simulation of the total pressure field for bianisotropic metasurfaces (left) and GSL metasurfaces (right) when $\theta_{\rm t}=60^\circ$, $70^\circ$, and $ 80^\circ$. }
		\label{fig:Fig3}
\end{figure*}

A simple way to study the bianisotropic response of the proposed particle is by analyzing the scattering produced by the particle. The scattering of the particle can be expressed in terms of the scattering matrix as 
\begin{equation}
\begin{bmatrix}
p_{\rm s}^+    \\
p_{\rm s}^- 
\end{bmatrix}=
\begin{bmatrix}
r^+  &   t^-  \\
t^+    &   r^-
\end{bmatrix}\begin{bmatrix}
p_{\rm i}^+    \\
p_{\rm i}^- 
\end{bmatrix}\label{eq:Tx_BC}
\end{equation}
where $p_{\rm i}^\pm $ represent the amplitude of the forward and backward incident plane waves, $p_{\rm s}^\pm $ is the amplitude of the scattered fields at both sides of the particle, $t^\pm$ represent the local transmission coefficients, $r^\pm$ are the reflection coefficients (the relation between the scattering matrix and the impedance matrix is detailed in Section Methods). 
Figure \ref{fig:Fig2D} shows the transmission and reflection amplitudes and phases for a particle defined by $h_2=1.5$ mm,$w_2=1$ mm, $h_1=1$ mm $w_1=4$ mm, $w_{\rm a}=6$ mm, $w_{\rm b}=5$ mm, $w_{\rm c}=4$ mm, and $w_{\rm d}=3$ mm.  For lossless and reciprocal particles the transmission coefficients satisfy $ t^+= t^-=t$ and reflection coefficient $|t|^2+|r^\pm|^2=1$.
The analysis of Fig. \ref{fig:Fig2D} shows that only the phase of the reflection is different for opposite directions, and this reflection phase asymmetry is a clear signature of bianisotropy \cite{radi2014tailoring,sieck2017origins}.

	\subsection{Design of acoustic bianisotropic gradient metasurfaces}

To demonstrate the applicability of the proposed bianisotropic cell, in what follows,  we will design  refractive metasurfaces for steering a normal incident wave ($\theta_{\rm i}=0^\circ$) into a transmitted wave propagating at $\theta_{\rm t}$. 
For a perfect refractive metasurface (with energy efficiency $\eta=100\%$), all the incident energy is redirected to the desired direction. This condition, equivalent to energy conservation in the normal direction, requires the macroscopic transmission coefficient to satisfy $T=1/\sqrt{\cos\theta_{\rm t}}$. Imposing the boundary conditions dictated by Eq.(\ref{eq:ImpedanceMatrix}), we can calculate the value of the impedance matrix at each point of the metasurface as
\begin{eqnarray}
Z_{11}=j{Z_0}\cot(\Phi_x x)\\
Z_{12}=j\frac{Z_0}{\sqrt{\cos\theta_{\rm t}}}\frac{1}{\sin(\Phi_x x)}\\
Z_{22}=j\frac{Z_0}{\cos\theta_{\rm t}}\cot(\Phi_x x) 
\label{eq:TransmitarrayCondition}\end{eqnarray}
where $\Phi_x=k\sin\theta_{\rm t}$ is the phase gradient along the metasurface and $Z_0$ is the characteristic acoustic impedance of the background medium. The period of the metasurface can be calculated as $D=2\pi/\Phi_x$. Equation (\ref{eq:TransmitarrayCondition}) shows that $Z_{11}$ is not equal to $Z_{22}$, so the bianisotropic response is required. This asymmetric behavior can be achieved by using bianisotropic cells proposed in this work.

The first design presented in this work corresponds to  $\theta_{\rm t}=60^\circ$. In this case, the required values for the components of the impedance matrix are represented in the left panel of Fig. \ref{fig:Fig3D}. The operating frequency is chosen to be 3000 Hz that makes the period of the metasurface $D=13.2$ cm. We use 11 cells along the period for implementing the spatial dependent bianisotropic response, so the width of the unit cell is $w=D/11=12$ mm.   In the discretization process, we choose the cells to have the impedance values at $x_n=(n-0.375)w$, where $n$ denotes the index of the cell, to avoid points where  the ideal impedance matrix diverges.
For the design of the physical dimensions, a  genetic algorithm optimization is used to define $w_1$, $w_{\rm a}$, $w_{\rm b}$, $w_{\rm c}$ and $w_{\rm d}$ so that the calculated impedance matrix matches the theoretical requirements. The physical dimensions of the final design are summarized in the Supplementary Material.  More details of the genetic algorithm optimization process, such as stop criteria, repeatability and evolution of the cost function can be found in the section Methods. 
 From Fig. \ref{fig:Fig3B}, we can see that the required impedance matrix of the perfect metasurface is closely approximated by our unit cells. It should be noted that the metasurface is discretized and approximated with a finite number of cells, and the performance of the metasurface can be possibly enhanced by using a larger number of cells with better spatial resolution.

\begin{figure}
\includegraphics[width=0.85\linewidth]{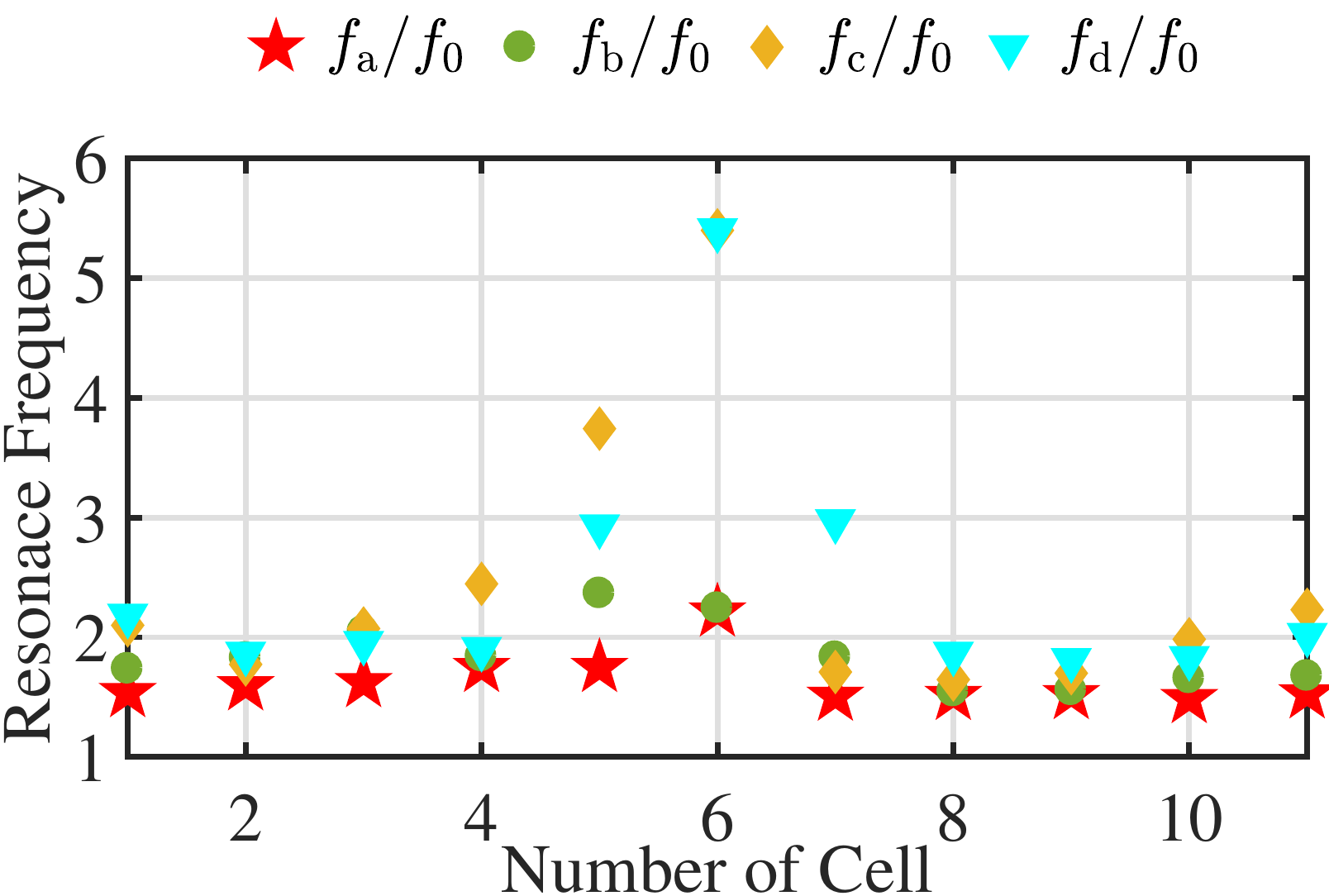}
\caption{Resonance frequency of the individual resonators of the scattering-free anomalous refractive metasurface designes for $\theta_{\rm i}=0^\circ$ and $\theta_{\rm t}=60^\circ$. All the resonators are working out of the resonant frequency.}\label{fig:Fig5}\end{figure}

Full-wave simulations are performed to verify our design. The real part of the simulated acoustic pressure  field for our first structure is represented in Fig. \ref{fig:Fig3A}, where nearly total energy transmission is observed. The simulated amplitude ratio $T$ achieved with our real structure is 1.365, as compared to the theoretically ideal value of 1.414, indicating that 93\%  of the incident energy is transmitted to the desired direction. This value is much higher than the theoretical upper limit of 89\% power transmission for conventional GSL based designs [see Fig. \ref{fig:Fig1B}]. With the purpose of comparison, we use a simulation of a discretized impedance-matched design based on the generalized Snell's law, confirming that in the conventional metasurfaces only 81\% of the input energy is transmitted in the desired direction, with the remainder going into reflection and other diffractive modes. Figure \ref{fig:Fig3A} shows the  comparison between the response of both designs, where we can clearly see the improvement obtained with the bianisotropic design. 

\begin{figure*}
\minipage{0.45\textwidth}
        \subfigure[]{\includegraphics[width=1\linewidth]{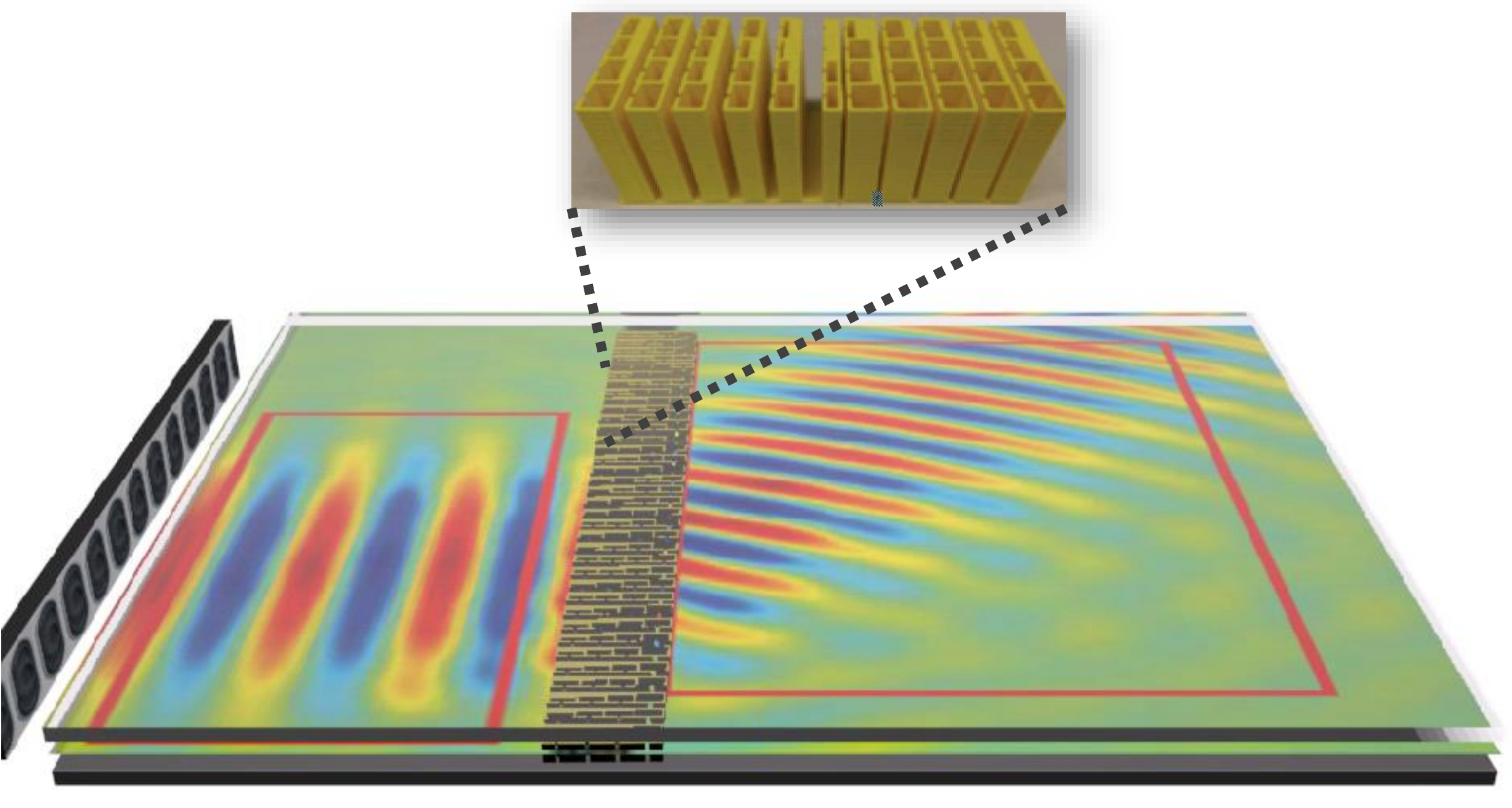}\label{fig:Fig4A}}
		\subfigure[]{\includegraphics[width=1\linewidth]{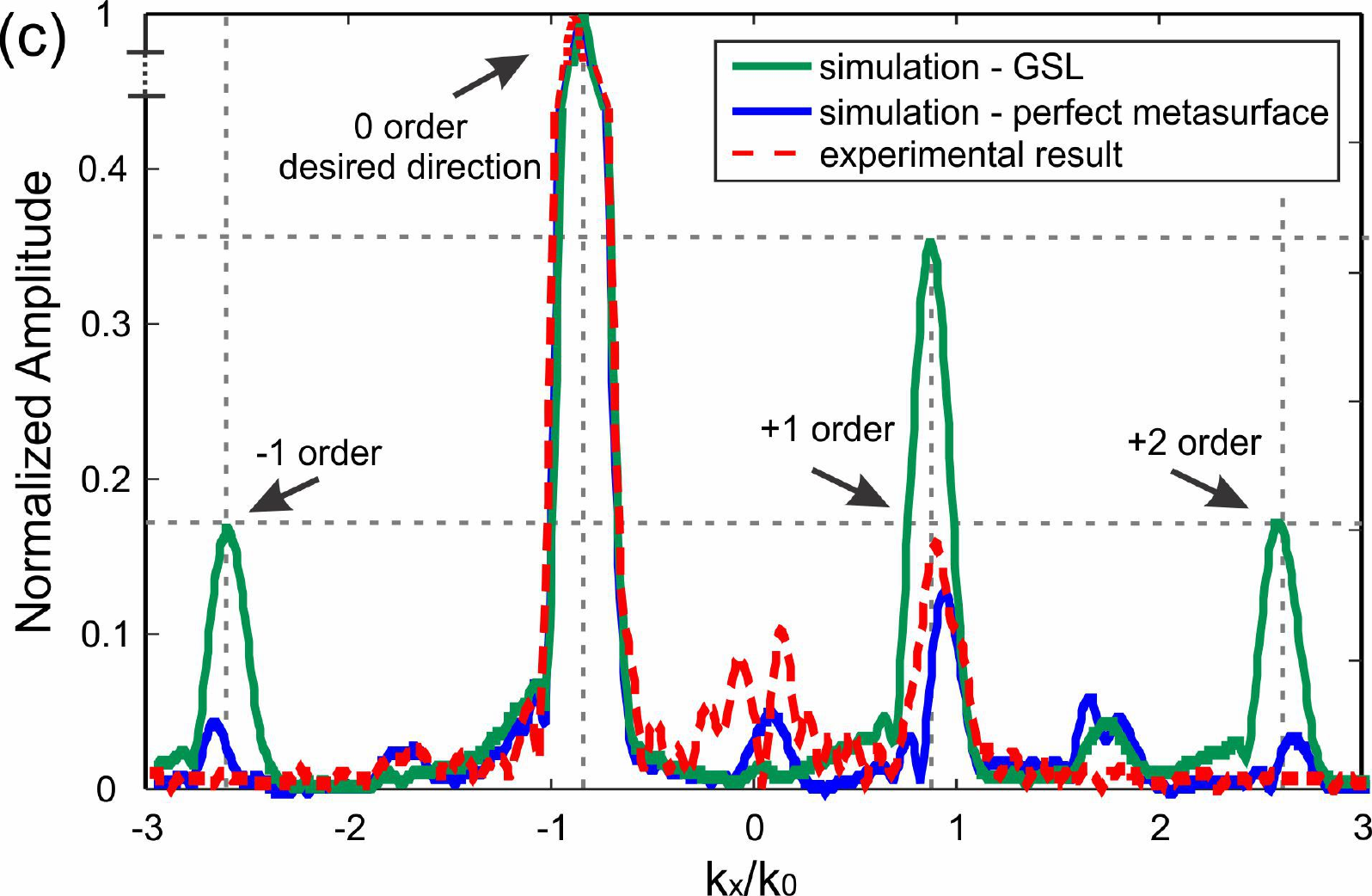}\label{fig:Fig4B}}
		\endminipage\hfill
		\minipage{0.32\textwidth}
\endminipage\hfill
		\minipage{0.5\textwidth}
		\subfigure[]{\includegraphics[width=1\linewidth]{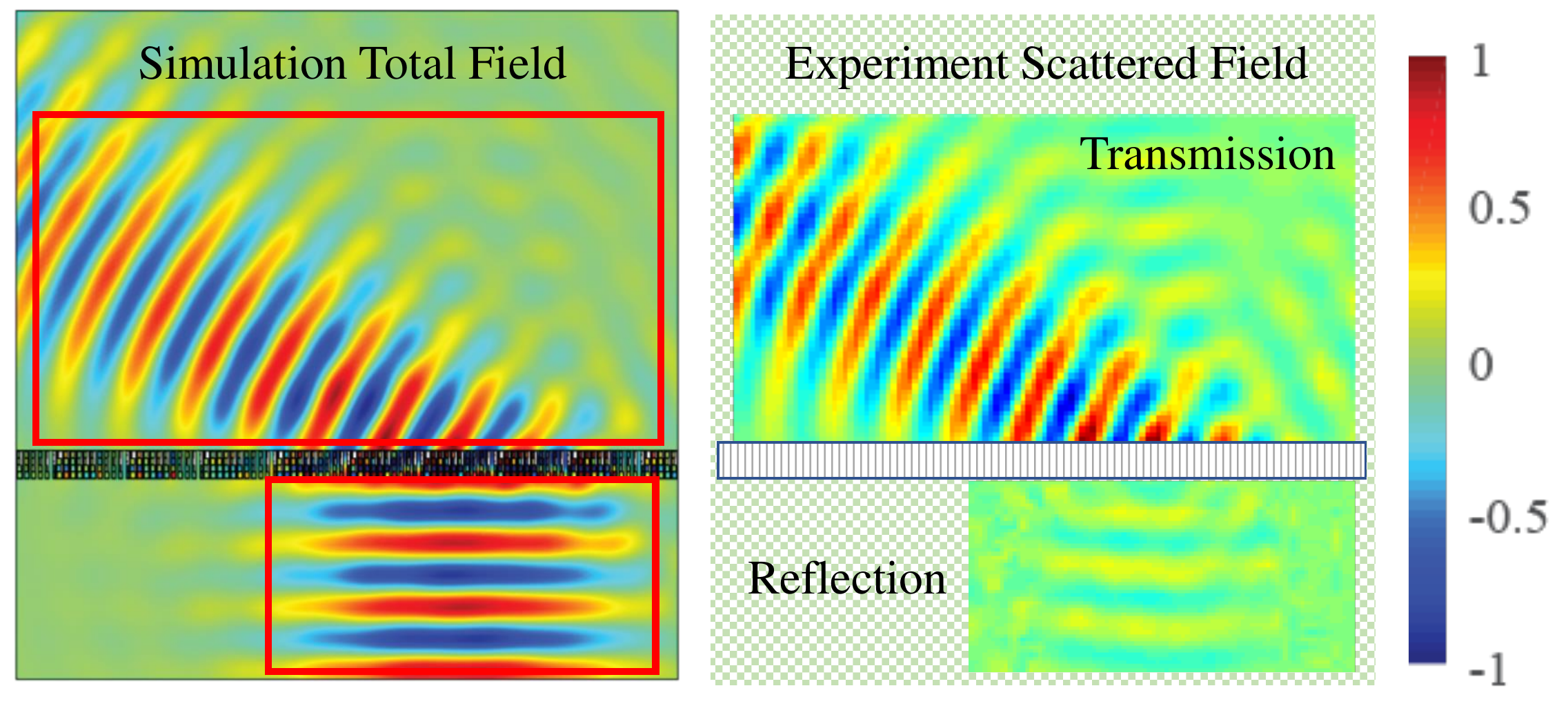}\label{fig:Fig4C}}
		\subfigure[]{\includegraphics[width=1\linewidth]{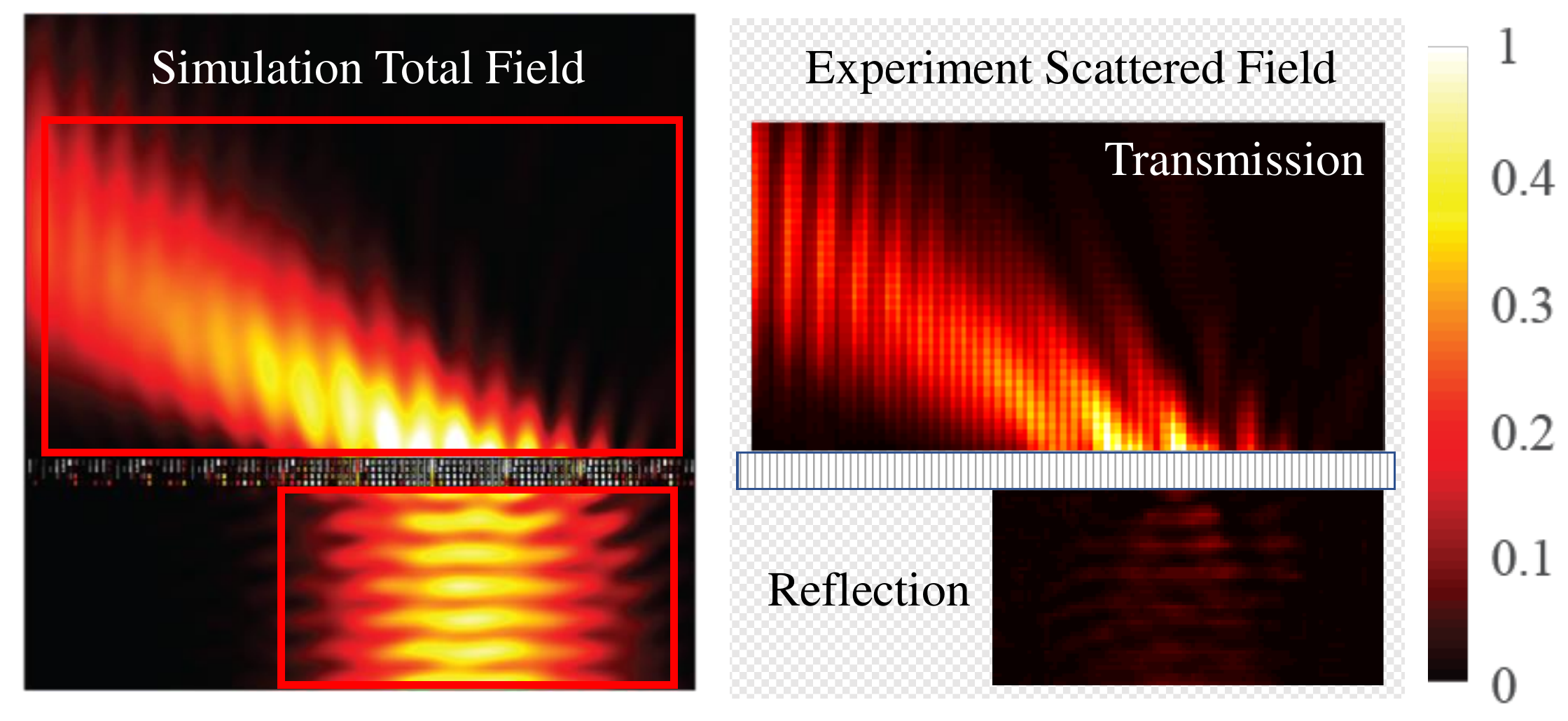}\label{fig:Fig4D}}
		\endminipage\hfill
		
\caption{(a) Schematic representation of the experimental setup and a period of the fabricated sample. (b) Comparison between the normalized scattering of the bianisotropic metasurface (experimental and numerical) and a GSL design. (c) Analysis of the real part (b) and magnitude (b) of the experimental pressure field and the comparison with the numerical simulations. }
\label{fig:Fig4}
	\end{figure*} 
 
Despite the piecewise constant and approximate realization of the theoretically ideal impedance profile, this practical structure nearly realizes perfect, lossless transmission of energy in the desired direction.  This shows that realistic structures can perform significantly better than conventional metasurfaces.  Critically, it also shows that good performance of a wavefront transformation metasurface does not require perfect realization of the ideal impedance profile.  A close and piecewise approximation will suffice in our design.

 Figure \ref{fig:Fig5} shows an analysis of the resonance frequency of each individual resonators. 
It is important to notice that none of the resonators is working near the resonance, so the design will be less sensitive to the losses than other resonant designs, as for example the three membrane proposal \cite{diaz2017acoustic}.  The performance of the design is also confirmed in simulation by considering viscous loss since it is the inherent loss of the structure which is inevitable in the experiments (Supplementary Note 3). In addition, due to the high resonance frequencies of the resonators, their size allows smaller width of the cells, i.e., it is easier to implement gradient metasurfaces with this topology.

To better show the large efficiency enhancement of the bianisotropic metasurface over conventional GSL-based designs, we designed another two cases with larger deflection angles, where the metasurfaces steer the incident beam to $\theta_{\rm t}=70^\circ$ and $\theta_{\rm t}=80^\circ$, respectively. For these two cases, the metasurfaces are sampled coarsely with only four cells within one period. The theoretical requirements (lines) and the achieved values (dots) of the impedance matrices for both cases are shown in Figs. \ref{fig:Fig3D} and \ref{fig:Fig3F}. Detailed dimensions and relative errors can be found in Supplementary Table 1 and 2. Fig. \ref{fig:Fig3F}  and Fig. \ref{fig:Fig3E}  show the simulated results of the bianisotropic designs (left) and the corresponding GSL-based designs (right) with ideal impedance matched cells and the same resolution. Energy efficiencies of the bianisotropic designs achieved 96\% and 91\% for $70^\circ$ and $80^\circ$ cases, whereas the corresponding numbers for GSL designs are 58\% and 35\%, respectively. Note that GSL-based designs are carried out by impedance matched cells with precise phase control, and the efficiency values are expected to be even lower for real structures.  We can see that even with such a coarse representation of the impedance profile and non-negligible relative error, the bianisotropic designs achieved much higher efficiency than the conventional ones. This offers huge advantage for practical realizations, especially in the high frequency or ultrasound range where fabrication capabilities are limited.

\subsection{Experimental measurements}

Measurements were carried out to characterize the design experimentally and confirm its scattering-free property.  As an example, we picked the $60^\circ$ case. The experimental setup and one period of the fabricated sample is shown in the Fig. \ref{fig:Fig4A}. The measured transmitted pressure field \ref{fig:Fig4C} and energy distribution  are shown in Fig. \ref{fig:Fig4D} are compared with the corresponding simulated fields. Good agreement between simulation and experiment is observed, and the small discrepancies can be attributed to fabrication errors and inevitable losses in the lab environment. The experimental results show that high-order diffractions are greatly suppressed and all the transmitted energy is concentrated in one direction. To confirm that our metasurface is reflection-free, the reflected field is also measured. The reflection caused by the metasurface is obtained by scanning the reflected region in the empty waveguide and the field with the metasurface, and then calculating the difference between the two measured fields. The result shows that only 2\% of the energy is reflected. 
To demonstrate the performance of the metasurface, the normalized energy distribution on each direction is further calculated by performing Fourier transform along the line right behind the metasurface, and the result is shown in Fig. \ref{fig:Fig4B}. The experimental result shows an excellent consistency with simulations, with most of the energy localized in the desired direction (zeroth order) and other diffraction modes are severely suppressed. 
The normalized energy distribution of a GSL based metasurface using impedance matched lossless effective medium computed from the same simulation shown in Fig. \ref{fig:Fig3A}, is also shown for comparison where high-order diffractions can be clearly observed. It should be noted that this number is calculated based on unit cells characterized by matched impedance and ideal refractive indices, and gives the performance limit of conventional designs. The bianisotropic metasurface proposed here therefore provides an alternative route of overcoming the power efficiency limitation and reduce the parasitic energy spread into undesired directions. 

\section{Discussion}

In summary, we design and experimentally demonstrate an acoustic metasurface cell that provides full control of the bianisotropic response and minimizes the implementation losses by ensuring that the individual resonators work below the resonant frequency.  The response of the cells, controlled by the physical sides of the four side-loaded resonators and the width of the channel, can be adjusted to provide  any scattering requirement. For a specific  asymmetric response, a carefully implemented GA optimization method calculates the physical dimensions of the unit cell. 

In addition, we have demonstrated the first design and realization of bianisotropic acoustic metasurfaces for scattering-free wavefront manipulations.
Three perfect metasurfaces for wavefront modulation (with deflection angle of $60^\circ$, $70^\circ$, $80^\circ$) are designed based on the theory. The performance is validated with numerical simulations, showing great advancement in energy efficiency (93\%, 96\%, 91\%) over conventional GSL-based designs (89\%, 58\%, 35\%), especially at large deflection angles. The scattering-free property of the bianisotropic metasurface is further verified experimentally. The designed metasurface is shown to be able to steer all the energy to the desired direction with almost no reflection or unwanted scattering.

We wish to emphasize that the proposed design scheme is not restricted to wave steering and can be readily extended to other applications. For example, similar “perfect” metasurfaces can be designed to achieve sound focusing without scattering, acoustic skin cloaking with low energy dissipation, and arbitrary acoustic field generation with high energy efficiency, among many others. In general, by considering non-local near field coupling and allowing the most general form of the cell’s impedance matrix, it will be possible to overcome the efficiency drawbacks in the existing metasurface designs. Also, since our bianisotropic design approach performs well even with a very coarse approximation of the continuous impedance profile, it offers great advantage in the ease of fabrication, especially for applications requiring a complicated field distribution, or extension to high frequency ranges. We would also like to point out that the proposed structure is not unique for realizing bianisotropic impedances, therefore improved lower loss structures are an important subsequent step towards applications. We believe that the bianisotropic metasurface concepts can largely expand the family of acoustic metasurfaces and open up new sound manipulation capabilities based on the versatile platform that can offer.

\section*{Methods}

\subsection*{Transfer matrix of the bianisotropic unit cell}
An analytic expression of the transfer function of the unit cells is developed to facilitate the design of the wavefront transformation metasurface. The geometry of a unit cell is shown in Fig. \ref{fig:Fig2A}, where $h_1$ is the thickness of the shell, $h_2$ is the width of the neck, $h_3$ is the length of the cavity, $w$ is the height of the unit cell, and $w_1$ and $w_2$ are the height of the channel and neck, respectively. The height of each individual Helmholtz resonator, $w_{\rm a,b,c,d}$, can be different as asymmetric geometry of the unit cell is required by the bianisotropic metasurface.

The relationship for the pressure and volume velocity of the incident and transmitted waves can be expressed as:
\begin{equation}
\begin{bmatrix}
p^{\rm +}    \\
\hat{n} \cdot\vec{u}^{\rm +}
\end{bmatrix}=
 \begin{bmatrix}
	M_{11}  &  M_{12}  \\
	M_{21}   & M_{22}
	\end{bmatrix}\begin{bmatrix}
 p^{\rm -}  \\
\hat{n} \cdot\vec{u}^{\rm -}
\end{bmatrix}\label{eq:TxMatrix}
\end{equation}
where $\vec{u}^{\pm}=w \ \vec{v}^{\pm}$, and $\left[M\right]$ is the total transfer matrix that can be written as:
\begin{equation}
\begin{split}
\left[M\right]=&\left[M_{\rm in} \right]\left[N_0\right]\left[M_{\rm a}\right]\left[N_0\right]\\
&\left[M_{\rm b}\right]\left[N_0\right]\left[M_{\rm c}\right]\left[N_0\right]\left[M_{\rm d}\right]\left[N_0\right]\left[M_{\rm out}\right].
\end{split}\label{eq:M}
\end{equation}
Here $\left[M_{\rm a}\right]$ through $\left[M_{\rm d}\right]$ are the transfer matrix of the individual Helmholtz resonator cell, and $\left[N_0\right]$ is the transfer matrix relating the Helmholtz resonator cells and the waveguide. The individual transfer matrix of the Helmholtz resonator cells A, B, C and D can be tuned by adjusting the geometries. The transfer matrices of the Helmholtz resonator cells (for example, cell A) and $N_0$ can be written as:
\begin{equation}
   \left[M_{\rm a}\right]=\begin{bmatrix}
	\frac{2-\alpha_{\rm a}}{2}  &  \frac{-\alpha_{\rm a}}{2}   \\
	\frac{\alpha_{\rm a}}{2}     &   \frac{2+\alpha_{\rm a}}{2} 
\end{bmatrix},\label{eq:Ma}
\end{equation}
and
\begin{equation}
   \left[N_0\right]=\begin{bmatrix}
	e^{jkh_1}  &  0  \\
	0    &   e^{-jkh_1}
	\end{bmatrix}.
\end{equation}
Here $\alpha_{\rm a} =R_{w1}/Z_{\rm a}$ and $R_{w1}=\rho _0c_0/w_1$ is the acoustic impedance of the straight channel, $Z_{\rm a}$ is the acoustic impedance of the Helmholtz resonator A. The same approach can be applied to the resonators B, C, and D.

The detailed derivation of $Z_{\rm a}$ is given in \cite{li2016theory}, and is directly given here for brevity: 
    \begin{equation}
    Z_{\rm a}=Z_{\rm n}\frac{Z_{\rm c}+jZ_{\rm n}\tan(kw_2)}{Z_{\rm n}+jZ_{\rm c}\tan(kw_2)}+j\Im{(Z_d)}.\label{eq:Zh}
    \end{equation}
Here $Z_{\rm n}=\rho _0c_0/h_2$ and $Z_{\rm c}$ are the acoustic impedance of the neck and the cavity of the Helmholtz resonator, respectively. Im($Z_{\rm d}$) is the radiation impedance between the neck and the straight channel and is expressed as:
\begin{equation}
  \begin{split}
    Z_{\rm d}=&\frac{\rho_0c_0}{w_1h_2^2}
    \frac{1-e^{-jkh_2}-jkh_2}{k^2}\\
    +&\frac{\rho_0c_0}{w_1h_2^2}\sum_{n=1} {\frac{1-e^{-jk_{zn}^\prime h_2}-jk_{zn}^\prime h_2}{{k_{zn}^\prime}^3}}
    \end{split}\label{eq:Zd}
\end{equation}
with $k_{zn}^\prime=\sqrt[]{k^2-{k_{xn}^\prime}^2}$ and  ${k_{xn}^\prime}=n \pi /w_1$. The acoustic  impedance of the cavity $Z_{\rm c}$ is given by:    
    \begin{equation}
     Z_{\rm c}=\sum_n\rho_0c_0\frac{k(1+e^{2jk''_{xn}w_3})\Phi _n^2}{k''_{xn}h_3(1-e^{2jk''_{xn}w_3})}. \label{eq:Zc}
    \end{equation}
where $\Phi _n=\sqrt[]{2- \delta _n}\cos (n \pi /2) {\rm sinc}(n \pi h_2/2 h_3)$ and $k''_{xn}=\sqrt[]{k^2-(n \pi /h_3)^2}$.

The transfer matrices of $\left[M_{\rm in}\right]$ and $\left[M_{\rm out}\right]$ are expressed as: 
\begin{equation}
   \left[M_{\rm in}\right]=\begin{bmatrix}
	\frac{1}{2}  &  \frac{R_{w1}}{2}   \\
	\frac{1}{2}  & -\frac{R_{w1}}{2} 
\end{bmatrix},
\end{equation}
and
\begin{equation}
   \left[M_{\rm out}\right]=\begin{bmatrix}
	1  &  1   \\
	\frac{1}{R_{w1}}   & -\frac{1}{R_{w1}}
\end{bmatrix}.
\end{equation}

By inserting Eqs.~(\ref{eq:Zh}-\ref{eq:Zc})  into Eq.~(\ref{eq:Ma}), the transfer matrix of an individual Helmholtz resonator unit can be obtained, which can further be combined with Eq.~(\ref{eq:M}) to compute the total transfer matrix. It can be seen that the total transfer matrix can be tuned by adjusting the geometrical values of the unit cell. In our design, $w$, $w_2$, $h_1$, $h_2$, $h_3$ are fixed, the heights of the Helmholtz resonator cells $w_{\rm a,b,c,d}$ and channel $w_1$ are put in the genetic algorithm to for the computation of the optimized structure.

Once the transfer matrix has been calculated, we can directly calculate the corresponding impedance matrix as
\begin{equation}
   \begin{bmatrix}
	Z_{11}  &  Z_{12}  \\
	Z_{21}   &   Z_{22}
	\end{bmatrix}= w \begin{bmatrix}
	\frac{M_{11}}{M_{21}}  & \frac{M_{11}M_{22}-M_{21}M_{12}}{M_{21}}  \\
	\frac{1}{M_{21}}  &  \frac{M_{22}}{M_{21}}
	\end{bmatrix}.
\end{equation}
These expressions have been used for calculating the actual impedance values in Figs. \ref{fig:Fig3B}, \ref{fig:Fig3D}, and \ref{fig:Fig3F}. Also, we can calculate the scattering matrix as
\begin{gather}
  \begin{bmatrix}
	 r^+ &   t^-  \\
    t^+    &    r^-
	\end{bmatrix}= \nonumber \\ 
  \begin{bmatrix}
	\frac{(Z_{11}-Z_0)(Z_{22}+Z_0)-Z_{21}Z_{12}}{\Delta Z}  &   \frac{2Z_{12}Z_0}{\Delta Z}  \\
  \frac{2Z_{21}Z_0}{\Delta Z}     &  \frac{(Z_{11}+Z_0)(Z_{22}-Z_0)-Z_{21}Z_{12}}{\Delta Z} 
	\end{bmatrix},
\end{gather}
where $\Delta Z=(Z_{11}+Z_0)(Z_{22}+Z_0)-Z_{21}Z_{12}$. This equation has been used for calculating the scattering coefficients represented in Fig. \ref{fig:Fig2}. A comparison between this  method and numerical simulations is presented in Supplementary Note 4.

\subsection*{Genetic algorithm}

In the optimization process, we used a genetic algorithm (GA) with continuous variables to find the optimized parameter for the resonators. The population size is 10 and the mutation rate is 0.2. We kept half of the genes for every generation and the best one does not mutate. There is no crossover in the optimization process. The optimization stops after 1500 generations. The algorithm is run 50 times for each cell to find the best match. In the design of $70^\circ$ and $80^\circ$ refraction, we used COMSOL Livelink with MATLAB to calculate the structures’ impedance matrices. More details about the convergence of the optimization process summarized in Supplementary Note 5.

\subsection*{Theoretical requirements for a scattering-free metasurface}
In the theoretical derivation of the impedance profile in Sec. II, we consider the following incident and transmitted pressure fields
\begin{equation}
p^+(x,y)=p_0e^{-jk(\sin \theta_{\rm i} x+\cos\theta_{\rm i} y)},
\end{equation}
and
\begin{equation}
p^-(x,y)=Tp_0e^{-jk(\sin \theta_{\rm t} x+\cos\theta_{\rm t} y)},
\end{equation}
where $p_0$ is the amplitude of the incident plane wave, $T$ is the transmission coefficient, and $\theta_{\rm i,t}$ are the angles of incidence and refraction. The velocity fields can be written as
\begin{equation}
\vec{v}^+(x,y)=\frac{p^+(x,y)}{Z_0}\left[\sin\theta_{\rm i} \hat{x}+\cos\theta_{\rm i} \hat{y}\right]
\end{equation}
and
\begin{equation}
\vec{v}^-(x,y)=\frac{p^-(x,y)}{Z_0}\left[\sin\theta_{\rm t} \hat{x}+\cos\theta_{\rm t} \hat{y}\right].
\end{equation}

For optimal performance of the metasurface, all the incident energy has to be redirected to the desired direction by a scattering-free metasurface. This condition, equivalent to energy conservation in the normal direction in all the point of the metasurface, $\hat{n}\cdot\vec{I}^+(x,0)= \hat{n}\cdot \vec{I}^-(x,0) $. Therefore, the required amplitude ratio of the transmitted wave and incident wave is given by
\begin{equation}
T=\sqrt[]{\frac{\cos \theta_{\rm i} }{\cos \theta_{\rm t}}}. \label{eq:amplitude}
\end{equation}
Expanding Eq.~(\ref{eq:ImpedanceMatrix}) with the assumed incident and transmitted fields, simplifying with the lossless and passive assumptions, $Z_{ij}=jX_{ij}$, and defining $\Phi_x =k(\sin \theta_{\rm i} -\sin \theta_{\rm t})$, the following relations are obtained by equating both the real and imaginary parts in the equation:
 \begin{align}
1&=T\frac{\cos\theta_{\rm t}}{Z_0}\sin(\Phi_x x) X_{12}\\
0&=\frac{\cos\theta_{\rm i}}{Z_0}X_{11}-T\frac{\cos\theta_{\rm t}}{Z_0}\cos(\Phi_x x) X_{12}\\
T \cos(\Phi_x x)&=T\frac{\cos\theta_{\rm t}}{Z_0}\sin(\Phi_x x) X_{22}\\
T \sin(\Phi_x x)&=\frac{\cos\theta_{\rm i}}{Z_0}X_{12}-T\frac{\cos\theta_{\rm t}}{Z_0}\cos(\Phi_x x) X_{22} \label{eq:system}
\end{align}
Putting the energy constraint shown in Eq.~(\ref{eq:amplitude})  into Eq.~(\ref{eq:system}), all the components of the impedance matrix can be obtained, yielding Eq.~(\ref{eq:TransmitarrayCondition}).
 
\subsection*{Numerical simulations}
The full wave simulations based on finite element analysis (FEA) are performed using COMSOL Multiphysics Pressure Acoustics module, where a spatially modulated Gaussian wave is incident normally on the metasurface. Perfectly matched layers (PMLs) are adopted to reduce the reflection on the boundaries. The loss in the air is modeled by the viscous fluid model in the Pressure Acoustic Module in COMSOL, with dynamic viscosity of $1.82\times 10^{-5}$ Pa$\cdot$s and bulk viscosity of $5.46\times 10^{-2}$ Pa$\cdot$s.

\subsection*{Experimental apparatus}
The samples were fabricated with fused deposition modeling (FDM) 3D printing. The printed material is acrylonitrile butadiene styrene (ABS) plastic with density of 1180 $\rm{kg/m^3}$ and speed of sound 2700 m/s, making the characteristic impedance much larger than that of air, and the walls are therefore considered to be acoustically rigid. The fabricated metasurface consists of 9 periods, and is secured in a two-dimensional waveguide for the measurement. A loudspeaker array with 28 speakers sends a Gaussian modulated beam normally to the metasurface and the transmitted field is scanned using a moving microphone with a step of 2 cm \cite{li2012design}. The acoustic field at each spot is then calculated using inverse Fourier Transform. The overall scanned area is 114 cm by 60 cm and the signal at each position is averaged out of four measurements to minimize noise.

\section*{Acknowledgments}
This work was supported by the Multidisciplinary University Research Initiative grant from the Office of Naval Research (N00014-13-1-0631) and in part by the Academy of Finland (projects 13287894 and 13309421).

\section*{Author contributions}
J.L., C.S., A.D. performed theoretical analysis and numerical simulations. J.L. and C.S. conducted the experiments and processed the data. All authors contributed to analyzing the data and preparing the manuscript. S.A.C. and S.A.T. supervised the study.

\bibliographystyle{aipauth4-1}
\bibliography{references}
\bibliographystyle{unsrt}

\end{document}